# Evidence of Topological Two-dimensional Metallic Surface States in Thin Bismuth Nanoribbons

Wei Ning,[†] Fengyu Kong,[†] Chuanying Xi,[†] David Graf,[‡] Haifeng Du,[†] Yuyan Han,[†] Jiyong Yang,[†] Kun Yang,[‡] Mingliang Tian,[†]* and Yuheng Zhang[†]

[†]High Magnetic Field Laboratory, Chinese Academy of Sciences, Hefei 230031, Anhui, P. R. China, and [‡]National High Magnetic Field Laboratory, Florida State University, Tallahassee, Florida 32306-4005, USA

* Address corresponding to tianml@hmfl.ac.cn

**Abstract**

Understanding of the exotic quantum phenomena in bulk bismuth beyond its ultraquantum limit still remains controversial and gives rise to a renewed interest. The focus of the issues is whether these quantum properties have a conventional bulk nature or just the surface effect due to the significant spin-orbital interaction and in relation to the Bi-based topological insulators. Here, we present angular-dependent magnetoresistance (AMR) measurements on single-crystal bismuth nanoribbons of different thickness with magnetic fields up to 31 T. In thin nanoribbons with thickness of ~40 nm, a two-fold rational symmetry of the low field AMR spectra and two sets of 1/2-shifted (*i.e.* $\gamma$=1/2) Shubnikov-de Haas (SdH) quantum oscillations with exact two- dimensional (2D) character were obtained. However, when the thickness of the ribbon increases, a 3D bulk-like SdH oscillations with $\gamma=0$ and a four-fold rotational symmetry of the AMR spectra appear. These results provided unambiguous transport evidence of the topological 2D metallic surface states in thinner nanoribbons with an insulating bulk. Our observations provide a promising pathway to understand the quantum phenomena in Bi arising from the surface states.

Semimetallic bismuth is a fascinating material with low carrier density, small effective mass and long mean free path, which distinguishes it from other metals as particularly suitable for studying quantum phenomena.[1-4] Recently, the surface states (SS) of Bi attracted a renewed interest due to its strong spin-orbital interaction[5] and in relation to the Bi-based topological insulators ($Bi_{1-x}Sb_x$, $Bi_2Se_3$, $Bi_2Te_3$).[6,7] It was found that the surface of an ultrathin Bi bilayers is quite distinct from the bulk semimetal but can be viewed as a quasi-two-dimensional (2D) metal with unique spin properties.[5] In fact, the SS on "cleaved" semi-infinite Bi have been found to show high electron density and large spin-orbit splitting due to the loss of the inversion symmetry.[8,9] Such metallic SS has, indeed, been confirmed by angle-resolved photoemission spectroscopy (ARPES)[10-12] and showed great interest on potential application in spintronics and the related fields. Unfortunately, the nature of the 2D SS in Bi remains inconclusive, especially in Bi with confined geometries, and there were no clear evidence of a topological 2D SS provided by angular-dependent Shubnikov-de Hass (SdH) quantum oscillations although some previous transport measurements were reported in Bi ultrathin films[13-15] or quasi-1D nanowires.[16-18]

In this work, we report angular-dependent magnetoresistance (AMR) measurements on individual single-crystal Bi nanoribbons in a large magnetic field ($B$) up to 31 T. Due to large surface-to-volume ratio, the nanoribbon with perfect planar surface is an ideal system to distinguish the 2D surface character from transport properties. Indeed, in our experiment, we observed clear angular-dependent SdH oscillations with 2D character, where the SdH spectra at various *B*-orientations can be well scaled with its perpendicular component ($B_\perp$) of the applied magnetic fields. In addition, low field AMR data show only a two-fold rotational symmetry in ~ 40 nm thick nanoribbon, instead of the expected three- or six-fold symmetry in semimetallic bulk Bi if rotating magnetic field in the bisectrix and binary plane.[19] These features provide a *direct* indication of an ideal 2D character of electronic structure, where the bulk contribution to the total MR is almost negligible due to the possible insulating nature. The carrier density, effective mass and the electron mean free path estimated from the 2D SdH are consistent with those in topological insulator $Bi_2Te_3$. Meanwhile, we found that the 2D SdH oscillations shows a 1/2-shifted behavior (*i.e. γ*=1/2), suggesting the existence of Dirac electrons on the surface and indicating a probable origin of a topological nature. Surprisingly, when the thickness of the nanoribbons increases to about 120 nm, a four-fold rotational symmetry of the low field AMR superimposed on the two-fold rotational spectra emerges, and gradually become dominant with the increase of the temperatures. Meanwhile, the SdH oscillations in thick ribbon were not seen in perpendicular magnetic fields, but appear in the parallel fields with *γ* = 0, which is in contrast to the case in ~ 40 nm thinner ribbons, indicating a 3D character of bulk effect. These results perhaps provide an insight on the nature of the SS of Bi in confined geometry and also an alternative understanding of the previous observations of the quantum phenomena in Bi crystals,[3,20] where the 2D SS might be responsible for the additional oscillation peaks far beyond the ultraquantum limit.[21]

**RESULTS AND DISCUSSION**

Bi nanoribbons were synthesized by solvothermal method.[22,23] Figure 1a shows the image of the nanoribbons by transmission electron microscopy (TEM), where the majority of nanoribbons display a typical width of 100 ~ 350 nm, a thickness (t) of 30 ~ 50 nm and a length of several micrometers. Selected area electron diffraction pattern on a randomly selected nanoribbon confirm its single-crystal morphology with a rhombohedral bulk crystal structure and a preferential [110] growth direction (*i.e.*, in hexagonal unit cell) (Figure S1 in Supporting information), which is perpendicular to the trigonal direction [001]. Further identification showed that the surface of the nanoribbon is built by the trigonal and bisectrix plane, which is similar to the case in ref 2 and 24 but the length direction of the ribbon is along the bisectrix axis.

To make standard four-probe device on an individual Bi nanoribbon for transport measurement, the nanoribbons were firstly dispersed on a silicon substrate with a 1 μm thick $Si_3N_4$ insulating layer and then transferred into FEI NanoLab 600i SEM/FIB dual beam system for deposition of electrodes. Four platinum (Pt) strips were deposited onto the ribbons along the [110] direction as the contact electrodes (see the inset of Figure 1b). The AMR measurement at high magnetic field was performed using standard a.c. lock-in techniques with a He-3 cryostat and a DC-resistive magnet (~31 T) at the National High Magnetic

Field Laboratory (NHMFL) at Tallahassee.

Resistance as a function of temperature for three nanoribbons NR1 (t~40 nm), NR2 (t~50 nm) and NR3 (t~120 nm) are shown in Figure 1b, where the magnitude of resistance in each curve was normalized to its value at 300 K. All three curves showed a non-metallic behavior which are similar to the observation in Bi nanowire.[16] A clear plateau feature with decreasing temperature was seen in both NR1 and NR2 thin nanoribbons. Such a plateau is very similar to that commonly observed in the element-doped topological insulators with an enhanced insulating bulk state,[25-28] where the surface conductivity plays a crucial role in low temperatures. It was well known that, in low-dimensional bismuth nanostructures, when the thickness of films or diameters of nanowires are comparable to the Fermi wavelength (~30 nm),[1] a semimetal to semiconductor transition could occur due to the quantum size confinement effect.[13,29,30] Therefore, the overall behavior of the temperature dependent resistance in thin NR1 and NR2 nanoribbons might be a consequence of the competition between the reduction of bulk conductivity and the increase of surface conductivity. In 120 nm thick Bi ribbon, the sample shows large residual resistance ratio without well recognized plateau feature in low temperature. This behavior is unexpected and might be due to the fact that once the thickness of nanoribbons exceeds a certain threshold, states become available in the bulk at Fermi energy and surface electrons transfer into the bulk, resulting in a big reduction of surface electron density. In the next, we will present unambiguous evidence of existence of such metallic surface states in thin Bi nanoribbons.

Figure 2a shows the MR *versus B* aligned at different angles, $\theta$, measured at 0.4 K for sample NR1, where the $\theta$ is the angle between *B* and the unit vector normal to the surface of the nanoribbon, as shown in the inset of Figure 2a. The MR is positive in low field region and then transition to negative in high magnetic field when the field is aligned perpendicular to the surface of the ribbons. This negative feature of MR is gradually weakened with the increase of the tilted angle. These complex MR properties are also seen in Bi nanowires.[18,31] The most remarkable results are that additional *R*-oscillations superimposed on the background of the MR curves can be well resolved both in low-*B* region ($B \leq$ 9 T) and high-*B* region ($B > 9$ T). Figure 2b shows the MR in low *B*-region as a function of $1/B_\perp$ with $B_\perp = B\cos\theta$. In high *B*-region, we took the MR variations, $\Delta R$, as a function of $1/B_\perp$, by subtracting the fitted smooth background, as shown in Figure 2c (The details of the background subtraction can be found in Figure S2 and Figure S3 in the supporting information). It was noted that both oscillation spectra at different angles approximately show periodic behavior with $1/B_\perp$, indicative of the SdH oscillations. More importantly, we found that all the maximum (or minimum) of these oscillations appear at the same $B_\perp$ for all angles, providing a direct evidence of a 2D character of the electronic state. Such a 2D SdH behavior could have a surface nature but we cannot completely exclude the 2D bulk origin because the thickness, ~ 40 nm of the ribbon is on the order of the Fermi wave length.[25,32]

To further identify these 2D SdH oscillations are from the exact surface states but not the bulk, we have performed the low-field AMR measurements on nanoribbons of different thickness. Figure 3a show the AMR of nanoribbon NR2 at a fixed magnetic field, $B = 1$ T as a function of the rotating angle $\theta$ in three different planes, and the schematic of the rotating geometries are shown in the insets of Figure 3a. We noted that, no matter in what planes the magnetic field rotated, the $R$-$\theta$ spectra always display a two-fold rotational symmetry, which significantly deviates from our expectations of three- or six-fold rotational symmetry when rotating the magnetic fields in the binary-bisectrix plane (if the bulk contribution is not negligible). This is because the electronic structure of bulk Bi includes three rotationally symmetrical electron pockets slightly off the bisectrix-binary plane.[24] These results provide firm evidence that the 2D transport character observed in thin nanoribbons is from the 2D surface states, the carriers from bulk electronic bands have no contribution on the MR. The two-fold rotational symmetric spectra are just a consequence of the usual magnetic scattering due to the Lorentz force, *e.g.*, the AMR presents a minimum (or maximum) when $B$ is in the surface plane or parallel to the applied current (or perpendicular to plane/current). The disappearance of the bulk conductivity is probably because the bulk in thin nanoribbons is already insulating due to the size confinement.[13]

As a comparison, Figure 3c presents the $R$-$\theta$ spectra of sample NR3 with a thickness ~ 120 nm in three different planes. Except for the two-fold symmetric spectra when rotating *B* in the binary - bisectrix plane, a four-fold rotational symmetric spectrum emerges from the background of the two-fold *R-θ* spectra when rotating *B* in both trigonal-bisectrix and trigonal-binary planes. When raising the temperature from below, the amplitude of the two-fold AMR decreases rapidly, but the four-fold AMR gradually become dominant at *T* >125 K as shown in Figure 3d. In contrast, for thin nanoribbon NR2, only two-fold AMR was seen with increasing temperature and the amplitude of the spectra decreases rapidly with the increase of temperature, as shown in Figure 3b. These comparative data clearly demonstrated that the two-fold symmetry of the AMR in Bi nanoribbons is from the 2D SS and dominates the MR in low temperatures but the four-fold symmetric AMR observed by rotating-*B* in both trigonal-binary and trigonal-bisectrix planes results from the 3D bulk-like effect due to both contributions of hole and electron carriers, where the schematic of the Fermi surface of the hole and electron pockets in three planes were shown in Figure S4 of supporting information. Obviously, the four-fold rotational symmetry of AMR due to the one hole pocket and two electron pockets are highly expected if rotating *B* in both trigonal-bisectrix and trigonal-binary planes (the hole pocket is almost normal to the two electron ones). Further evidence of 3D bulk effect in sample NR3 was also obtained by the AMR measurement with magnetic field *B* up to 31 T, as shown in Figure 4a.

We found that the MR oscillations were only observed within ±10° when *B* is applied around the length direction of the nanoribbon, as shown in Figure 4b.

Figure 4c is the MR variations, $\Delta R$, as a function of $1/B_\perp$, the periodic behavior with $1/B$ can be observed when $\theta > 80°$. This feature is almost in opposite to the case in ~ 40 nm thinner ribbons, where the significant 2D SdH signal can be well resolved when $\theta < 50°$ (*i.e.*, with a larger component of $B_\perp = B\cos\theta$). Since the SdH spectra in 2D electronic system is not possible in parallel $B$, the observed SdH in ~120 nm thick ribbon near parallel $B$ is clearly attributed to the bulk contribution so that the SS was considerably smeared out by the bulk effect. The absence of oscillations under $B_\perp$ may be attributed to the special properties of wave function of the bulk states in this confined geometry. Since the thickness of NR3 is still comparable to the bulk Fermi wavelength, the wave function in the direction perpendicular to the nanoribbon are discrete confined states; such discreteness modifies density of states, possibly resulting in smaller (2D) Fermi wave vector in the ribbon plane. Moreover, these states may actually be Anderson localized in the ribbon due to lack of topological protection (unlike surface states), resulting in the insulating behavior (shown in Figure 1b) and lack of quantum oscillation under $B_\perp$ in NR3. On the other hand $B_{//}$ couples to the (discrete) wave functions in the perpendicular direction and affects their confinement energy; such coupling modifies density of states and can give rise to observed oscillation.

Figure 4d shows the plot of $1/B$ *versus* Landau index number $n$ for both samples NR1 and NR3. In a 2D metal, the SdH oscillations can be described by the Onsager relation $2\pi(n+\gamma) = S_F(\hbar/eB)$,[25] where the Landau level index $n$ is related to the extremal cross-section, $S_F$, of the Fermi surface with $S_F = \pi k_F^2$, and $k_F$ is the Fermi vector, $e$ is the electron charge, and $\hbar$ is Planck's constant divided by $2\pi$, and the phase factor $\gamma$ is equal to 0 for a regular electron gas or 1/2 for an ideal topological 2D Dirac fermions, respectively. From the linear fitting of the data in thinner NR1 as shown in Figure 4d, the slopes yield the frequency of the oscillations, $F = \frac{\hbar}{2\pi e} S_F$ =12.2 T and 44.6 T, with $k_F \sim$ 0.019 Å$^{-1}$ and 0.036 Å$^{-1}$, respectively. As the 2D surface carrier density $n_{2D}$ is related to $k_F$ by $n_{2D} = k_F^2/4\pi$, thus we obtain $n_{2D}$=2.8×10$^{11}$ cm$^{-2}$ and 1.1×10$^{12}$ cm$^{-2}$ for the surface carries, respectively. In addition, the linear extrapolation of the 1/$B$ ~ n relationship to 1/$B$ = 0 yields a finite intercept of $\gamma$ = 0.40 and 0.51 for low-field region SdH and high-field region SdH, respectively. They are very close to the ideal value of 0.5 for Dirac electrons, resulting in a 1/2-shifted SdH oscillations, which suggest the topological nature of the SS. The observation of two set of SdH oscillations in surface states implies the possible multiband electronic structures, such as multi-Dirac cones. Another possibility is that the two sets of 2D SdH origin from different surface states of the nanoribbons where the bottom surface contacts with the substrate and has different carrier density with the top surface. For sample NR3, the frequency $F$ of the oscillations is 16.3 T, with $k_F \sim$ 0.022 Å$^{-1}$, thus the period of SdH oscillations related to a 3D Fermi sphere give a carrier density of $n_{3D} = k_F^3/3\pi^2$ =3.5×10$^{17}$ cm$^{-3}$. This value is very close to the bulk carries density of 3×10$^{17}$ cm$^{-3}$. As a comparison, the intercept, $\gamma$, of NR3 is about 0.03, close to zero and suggesting to conventional Schrödinger spectrum.

This is also consistent with MR behavior at $\theta = 90°$ where only a slightly negative MR can be observed, suggesting usual localization for Schrödinger electrons. As comparison, a strong positive MR can be observed in sample NR1 which suggests antilocalization due to the strong spin-orbit coupling and in particular Dirac spectrum.

To extract more specific information on the surface states, we next analyzed, as an example, how the SdH amplitudes in the high B-regime vary with temperature. Figure 5a shows the $\Delta R$ vs. $1/B$ aligned perpendicular to nanoribbon at different temperatures. The oscillation amplitudes in $\Delta R$ decrease rapidly as $T$ increases from 0.4 K to 10 K. The temperature-dependent amplitude $\Delta R$ of the SdH oscillations can be described as $\Delta R(T)/\Delta R(0) = \lambda(T)/\sinh(\lambda(T))$ [32,33] with the thermal factor of $\lambda(T) = 2\pi^2 k_B T m^*/(\hbar eB)$, where $m^*$ is the effective mass of surface state carriers. By taking the oscillation amplitude of the peak $n = 2$, the effective mass of surface carrier, $m^*$ is extracted to be ~ 0.35 $m_0$ by performing a theoretical fit with above equation, as shown in Figure 5b, $m_0$ is the static electron mass. This value is much larger than the average effective mass of the carriers for bulk Bi (~ 0.065 $m_0$), but is on the order of the effective mass of the surface-state carriers in topological insulator, $Bi_2Se_3$ or $Bi_2Te_3$.[25,26,32-35] As a result, the Fermi velocity is calculated as $V_F = \hbar k_F/m^* = 1.2 \times 10^5$ ms$^{-1}$, and the transport lifetime of the surface states, $\tau$ can be estimated by utilizing the Dingle relation. Since $\Delta R/R_0 \sim [\lambda(T)/\sinh\lambda(T)]e^{-D}$, where $D = 2\pi^2 E_F/\tau eBV_F^2$, $E_F = m^* V_F^2$, $R_0$ is the classical resistance at a zero field, the lifetime $\tau$ can be derived from the slope in Dingle plot by $\ln[(\Delta R/R_0)B\sinh\lambda(T)]\approx[2\pi^2 E_F/(\tau eV_F^2)]\times(1/B)$ (see Figure 5c). Hence, the surface state lifetime $\tau$ of carriers is estimated to be $1.17\times10^{-12}$ s, and thus we get an electron mean free path, $\ell=V_F\tau \sim 140$ nm and the surface mobility $\mu_s = e\tau/m^* = e\ell/\hbar k_F \sim 5840$ cm$^2$ V$^{-1}$ S$^{-1}$. Thus, the metallicity parameter, $k_F\ell$, is then estimated to be 50, which is very close to the value of ~ 66 obtained from the SS of high quality Bi$_2$Te$_3$ bulk crystals and nanowires.[25,35] Similar analyses of data for NR1 in the low-field regime and for NR3 have also been made and the results were shown in Figure S5 and S6 of supporting information. From the fitting, we get the effective carrier mass $m^*$ to be 0.21 $m_0$ of thin nanoribbon NR1 in low-field regime, which is smaller than that in the high-field regime. Meanwhile, the effective mass of sample NR3 is 0.19 $m_0$.

It is noted that all of the single-crystal nanoribbons measured in this work were naturally oxidized on the surface during the process of the nanofabrication, the survival of the ideal isotropic-like 2D surface states in thin Bi nanoribbons provide an indication that such surface states are rather robust to their environment which is quite in contrast to the normal SS in semiconductors or others. Meanwhile, the measured carrier density, mobility, electron mean free path and the effective mass of carriers in the SS of Bi nanoribbons are all on the order of those obtained in topological insulators, such as in Bi$_2$Se$_3$ or Bi$_2$Te$_3$ bulk or nanowires. Especially we observed 1/2-shifted SdH oscillations of the surface states which suggests the existence of Dirac electrons on the surface states. These results indicate that the surface states in Bi nanoribbons should be topologically nontrivial. Recently, transport and ARPES studies on Bi (111) bilayer have been made to address its topological nature and suggestive evidence for the existence of edge state states has been reported.[36-39] Thus, the observation of topological surface states in our thin nanoribbons might not be surprising which might be stabilized or formed by the size confinement effect in Bi nanoribbons. A recently work on Bi films also demonstrated that the films changed from topologically trivial to topological nontrivial when the thickness is thin enough.[40] However, to full understand how the nontrivial topology properties evolved from bulk crystal to thin films, further experimental studies are needed. A fact that, in thin Bi nanoribbons with a thickness less than 50 nm, the $R$-$\theta$ spectra always show a two-fold rotational symmetry regardless of the field orientation. If this 2D surface states of the nanoribbon is trivial, a six-fold rotational symmetry of the $R$-$\theta$ spectra with $B$-oriented in the binary-bisextrix plane will be expected because the trivial SS strongly depends on the bulk electronic structure, such as a hexagonal electron pocket around the $\Gamma$–point and six hole lobes along the $\Gamma$-M direction formed in the surface Brillouin zonein ultrathin Bi (111) or (110) bilayers.[10,11] Hence, the isotropic 2D nature of the SS in thin Bi ribbons is consistent with the expectation of the topological SS with a circular Fermi Surface of the Dirac cone. Additionally, our studies also suggested that the side surfaces of ribbons have the same topological properties as the top and bottom surfaces. For sample NR1, a weak antilocalization (WAL) with 2D characteristic can be identified at the low-field range. Further analysis of the 2D WAL effect suggests that the metallic surface states of our Bi nanoribbons (top, bottom and side surfaces together) forms a continuous 2D spin-orbit metal as expected for the surface states of topological insulator. Further numerical calculations of band structures or ARPES measurements on an individual thin Bi nanoribbon might be very helpful for verification of the nature of the surface state.

**CONCLUSION**

In summary, we have performed AMR studies on Bi nanoribbons of different thickness with magnetic field up to 31 T. Through the AMR data, we have identified two sets of the SdH oscillations with 1/2-shifted in thinner nanoribbons due to the exact 2D metallic surface states in low temperatures, and also demonstrated that the 3D bulk states are insulating due to the size confinement effect. The surface carries density and effective mass obtained from the SdH oscillations are on the same order of those of the surface states in topological insulators ($Bi_{1-x}Sb_x$, $Bi_2Se_3$, $Bi_2Te_3$). Our experimental results indicate that the surface states in the nanoribbons might be topological protected. When the thickness of the nanoribbon is larger than 120 nm, the 3D bulk state appears and gradually dominates with the increase of temperatures.

METHODS

The Bi nanoribbons were synthesized by solvothermal method. In a typical process, 0.15 g analytical grade sodium bismuthate ($NaBiO3.2H2O$) dissolved in 30 ml glycerol, stirred vigorously for 30 min, and then the solution was transferred to Teflon-lined stainless steel autoclave with a capacity of 50 ml. The solution was bubbled with a flow of pure nitrogen gas for 10 min, before the autoclave was sealed and maintained at 200℃ for 24 h. After reaction, the resulting black solid product was collected by filtration, washed with ethanol to remove all of impurities. The structural characteristics of Bi nanoribbons were investigated using an JEOL—2011 TEM.

*Supporting Information Available.* Experimental details for fabricating the Bi devices, detailed subtraction of the background to get the SdH oscillations, and the schematic of the Fermi surface of the hole and electron pockets in three planes for bismuth from the literature. This material is available free of charge *via* the Internet at http://pubs.acs.org.

*Conflict of Interest*: The authors declare no competing financial interest.

*Acknowledgement.* This work was supported by the National Key Basic Research of China, under Grant Nos. 2011CBA00111 and 2010CB923403; the National Nature Science Foundation of China, Grant No. U1332139, No. 11104280, No. 11174294, and No. 11104281. Dr. Tian also thanks the support of the Hundred Talents Program of the Chinese Academy of Science and NHMFL funded by NSF under Grant No. DMR-0654118, the State of Florida, and the U.S.

Department of Energy. KY was supported by NSF under Grant No. DMR-1004545.**REFERENCE AND NOTES**

1. Ashcroft, N. W.; Mermin, D. N. *Solid state physics* (Thomson Learning, Toronto, 1976), 1st ed.
2. Yang, F. Y.; Liu, K.; Hong, K., Reich, D. H., Searson, P. C. Chien, C. L. Large Magnetoresistance of Electrodeposited Single-crystal Bismuth Thin Films. *Science* **1999**, *284*, 1335-1337.
3. Behnia, K.; Balicas, L.; Kopeleivch, Y. Signatures of Electron Fractionalization in Ultraquantum Bismuth. *Science* **2007**, *317*, 1729-1731.
4. Li, L.; Checkelsky, J. G.; Hor, Y. S.; Uher, C.; Hebard, A. F.; Cava, R. J.; Ong, N. P. Phase Transitions of Dirac Electrons in Bismuth. *Science* **2008**, *321*, 547-550.
5. Hofmann, P. The Surfaces of Bismuth: Structural and Electronic Properties. *Prog. Surf. Sci.* **2006**, *81*, 191-245.
6. Hasan, M. Z.; Kane, C. L. Colloquium: Topological Insulators. *Rev. Mod. Phys.* **2010**, *82*, 3045-3067.
7. Qi, X. -L.; Zhang, S.-C. Topological Insulator and Superconductor. *Rev. Mod. Phys.* **2011**, *83*, 1057-1110.
8. Koroteev, Yu. M.; Bihlmayer, G.; Gayone, J. E.; Chulkov, E. V.; Blügel, S.; Echenique, P. M.; Hofmann, Ph. Strong Spin-orbit Splitting on Bi Surfaces. *Phys. Rev. Lett.* **2004**, *93*, 046403.
9. Pascual, J. I.; Bihlmayer, G.; Koroteev, Yu. M.; Rust, H.-P.; Ceballos, G.; Hansmann, M.; Horn, K.; Chulkov, E. V.; Blügel, S.; Echenique, P. M.; *et al.* Role of Spin in Quasiparticle Interference. *Phys. Rev. Lett.* **2004**, *93*, 196802.
10. Ast, C. R.; Höchst, H. Electronic Structure of a Bismuth Bilayer. *Phys. Rev. B* **2003**, *67*, 113102.
11. Hirahara, T.; Nagao, T.; Matsuda, I.; Bihlmayer, G.; Chulkov, E.V.; Koroteev, Yu. M.; Echenique, P. M.; Saito, M.; Hasegawa, S. Role of Spin-orbit17

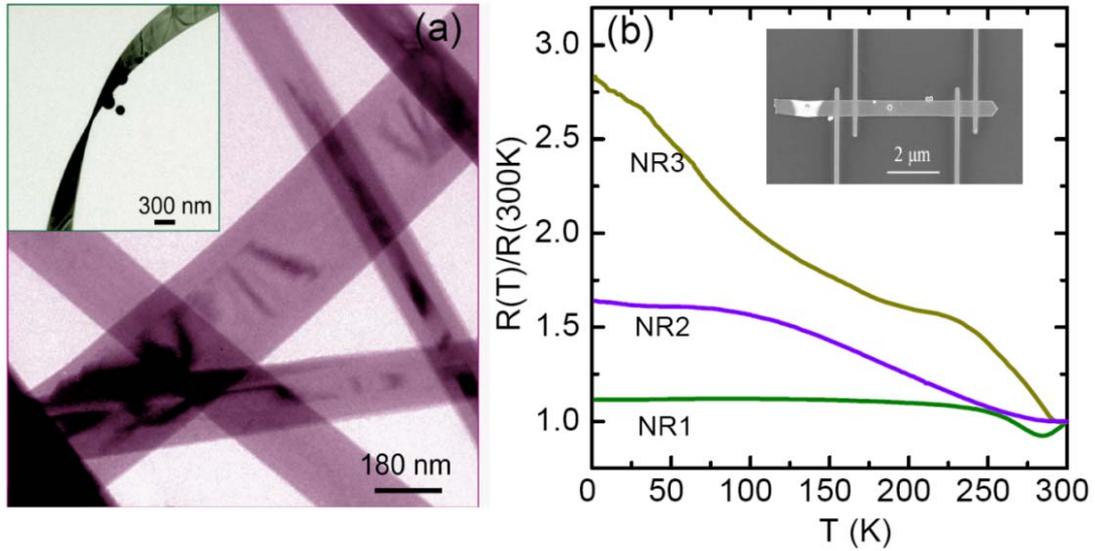

**Figure 1.** TEM image and temperature-dependent resistance of Bi nanoribbons. (**a**) transmission electron microscopy image of nanoribbons, and the inset shows a twisted ribbon. (**b**) The resistance as a function of temperature of three individual nanoribbons, NR1, NR2 and NR3 with thickness of 40 nm, 50nm and 120nm, respectively. The inset is a SEM image of a nanoribbon for four-probe transport measurements.

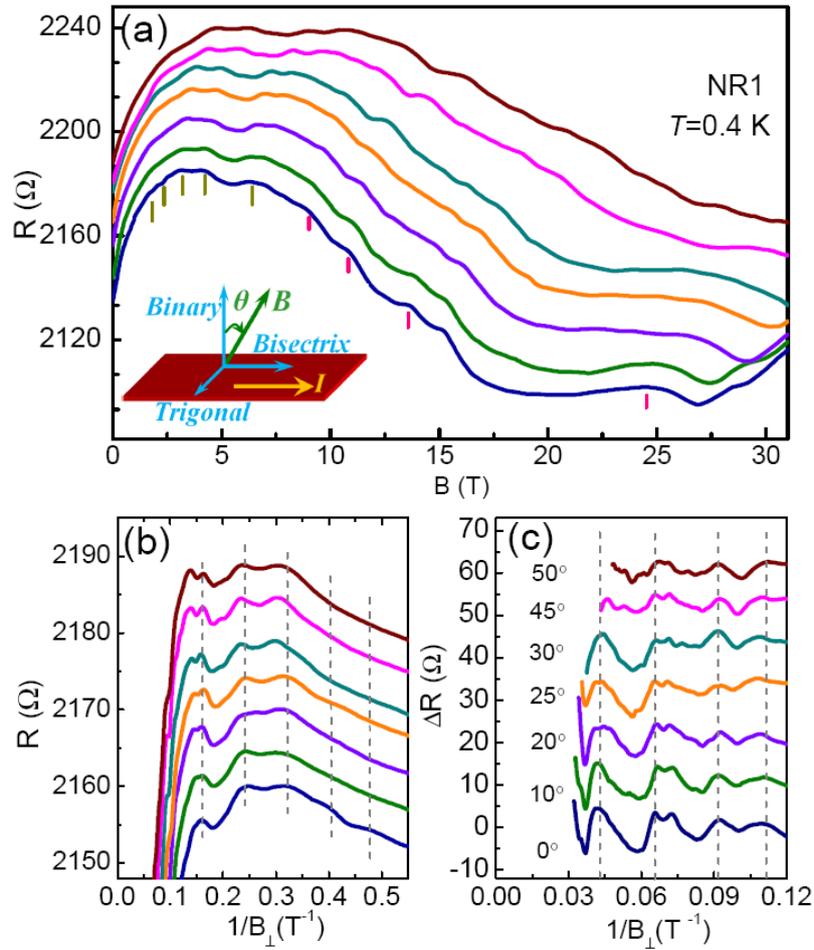

**Figure 2.** 2D SdH oscillations of conduction in ~ 40 nm Bi nanoribbon. (a) The resistance of sample NR1 (~40 nm) as a function of magnetic field at different tilted angle, $\theta$, at 0.4 K ($B$ rotates within binary and bisectrix plane). The inset is the schematic of $B$ orientation in different planes of nanoribbon. (b) The enlarged MR in low $B$-regime of $B \leq 9$ T as a function of $1/B_\perp$. (c) Amplitude of the resistance oscillations, $\Delta R$, in high $B$-regime ($B > 9$ T) *versus* $1/B_\perp$. Both oscillation spectra show periodic behavior with $1/B\cos\theta$, indicating a typical 2D character of electronic structure. Each curve was offset for clarity.

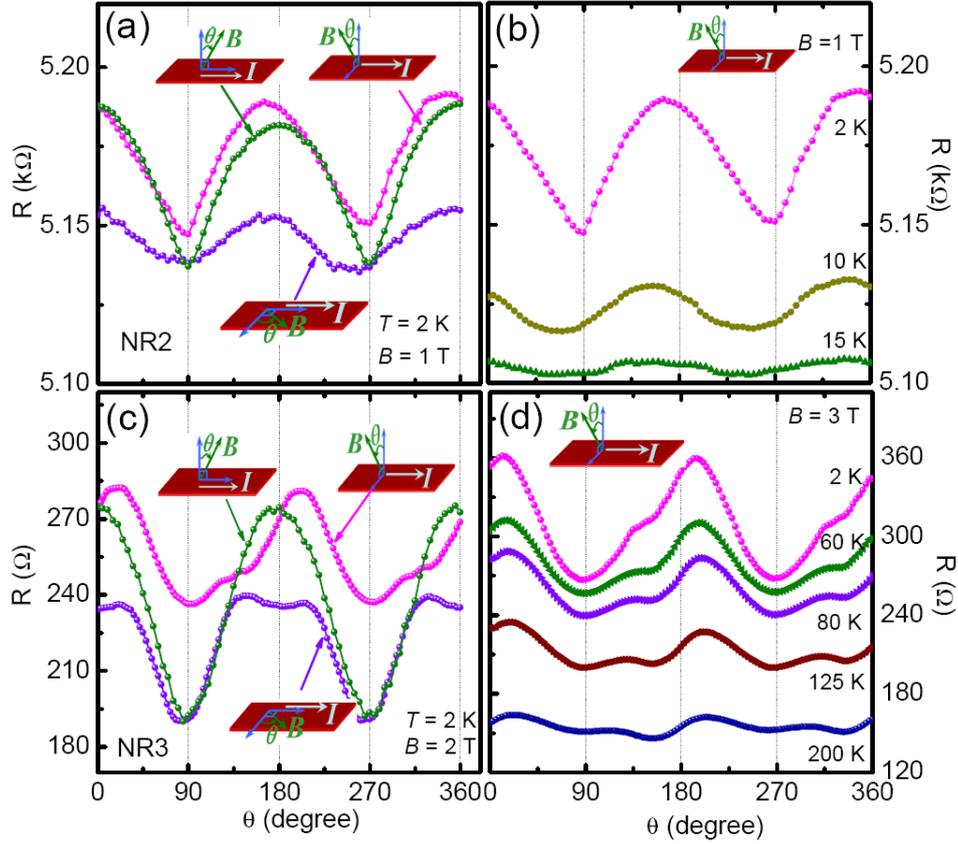

**Figure 3.** Low-field AMR of nanoribbons with different thickness. (a) $R$-$\theta$ spectra of sample NR2 (~ 50 nm) obtained at 2 K with an external magnetic field 1 T rotating in three different planes. (b) $R$-$\theta$ spectra of sample NR2 with rotating B within binary and trigonal plane at different temperatures. (c) $R$-$\theta$ spectra of sample NR3 (~120 nm) obtained at 2 K with an external magnetic field 2 T rotating in three different planes. (d) $R$-$\theta$ spectra with rotating B within binary and trigonal plane at 3 T measured at different temperatures.

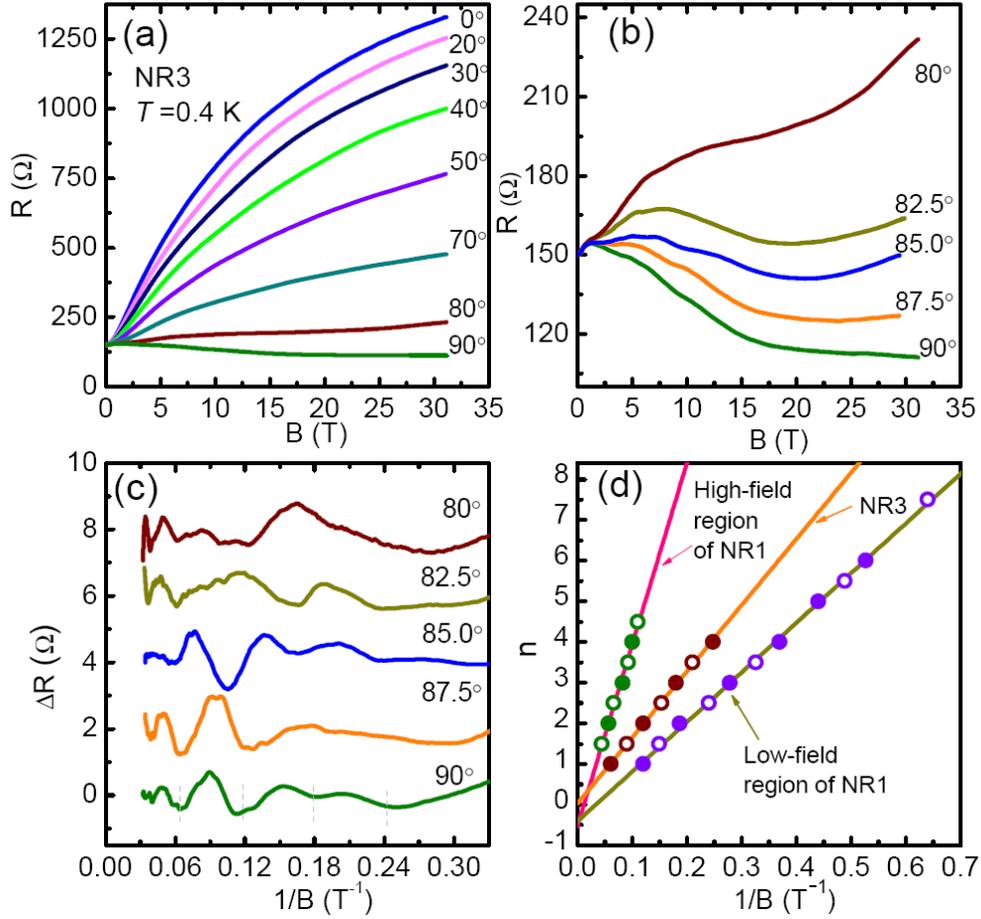

**Figure 4.** 3D SdH oscillations of conduction in ~ 120 nm thick Bi nanoribbon. (a) The resistance of sample NR3 (~120 nm) as a function of magnetic field at different tilted angle, $\theta$, at 0.4 K (B rotates within binary and bisectrix plane). (b) The enlarged MR as a function of magnetic field with $\theta$ between 80° to 90°. (c) Amplitude of the resistance oscillations $\Delta R$ versus $1/B_\perp$. Each curve was offset for clarity. (d) 1/B versus the Landau index number, $n$, for both sample NR1 and NR3. The dips and peaks of oscillations correspond to n and n+1/2 and are denoted by the open and filled circles, respectively. The solid lines correspond to the best fits to the experimental data.

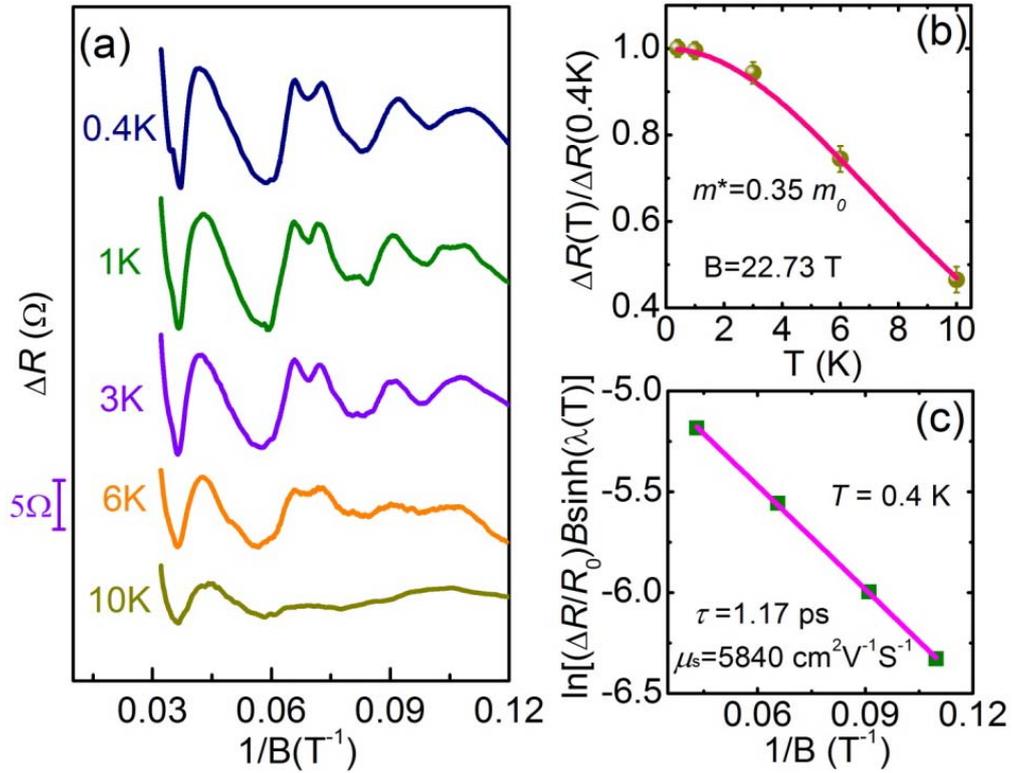

**Figure 5.** Temperature-dependent SdH oscillations for sample NR1 (~40 nm) in the high-field regime. (**a**) The amplitude of the SdH oscillations *versus* 1/*B* of under a perpendicular *B* at different *T*. (b) Temperature dependence of the scaled oscillation amplitudes at *B*=22.73 T. (**c**) Dingle plot of the SdH oscillations at 0.4 K.

**TOC Graphic:**

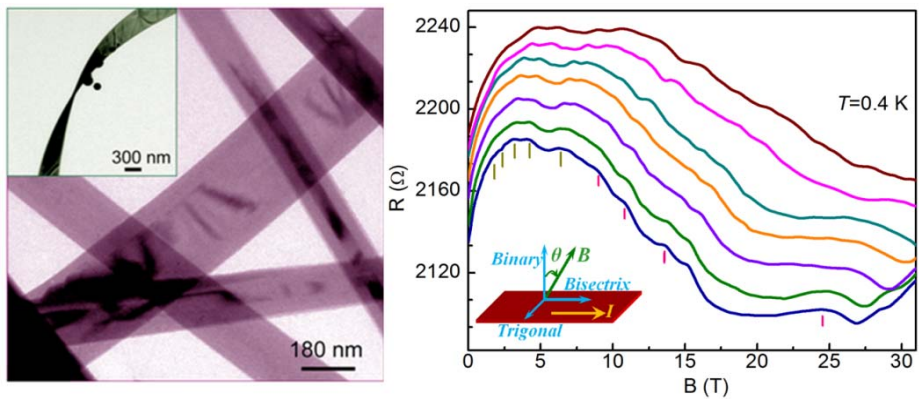

# Evidence of topological two-dimensional metallic surface states in thin bismuth nanoribbons

*Wei Ning,[†] Fengyu Kong,[†] Chuanying Xi,[†] David Graf,[‡] Haifeng Du,[†] Yuyan Han,[†] Jiyong Yang,[†] Kun Yang,[‡] Mingliang Tian,[†]\* and Yuheng Zhang[†]*

[†]High Magnetic Field Laboratory, Chinese Academy of Sciences, Hefei 230031, Anhui, P. R. China; Hefei National Laboratory for Physical Science at The Microscale, University of Science and Technology of China, Hefei 230026, People's Republic of China

[‡]National High Magnetic Field Laboratory, Florida State University, Tallahassee, Florida 32306-4005, USA

\* Address corresponding to tianml@hmfl.ac.cn

**I. A TEM image of an individual nanoribbon and the schematic of the Fermi surface of the hole and electron pockets with respect to the crystal structure of Bi.**

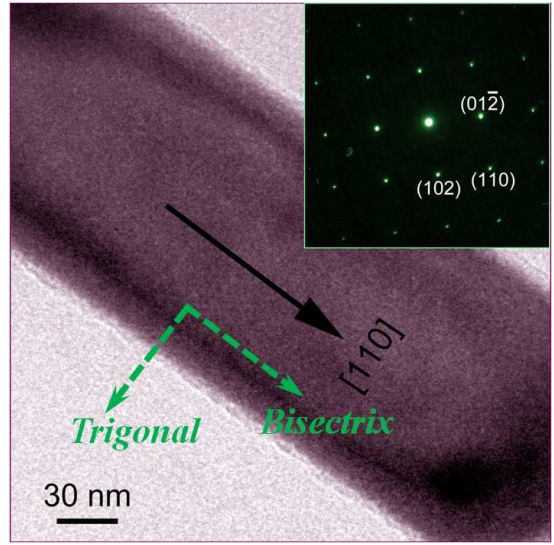

Figure S1. A TEM image of an individual nanoribbon. The growth direction of the nanoribbon is along (110), which is perpendicular to the trigonal direction [001].

**II. Experimental Details for Fabricating Bi devices**

To make standard four-probe devices on an individual Bi nanoribbon for transport measurement, the nanoribbons were firstly dispersed on a silicon substrate with a 1 mm thick Si3N4 insulating layer and then transferred into FEI NanoLab 600i SEM/FIB dual beam system for the deposition of electrodes. Four platinum (Pt) strips with width of 100 nm and a thickness of 150 nm were deposited onto the ribbons as the contact electrodes (the beam current is 7 pA with an excitation voltage of 30 kV). The maximum spreading distance of Pt-deposition along the nanowires was found to be less than 200 nm beyond the intended position through a profile analysis of the TEM energy dispersive X-ray study. Such a spreading of Pt-deposition has less effect on the transport measurement of the wire due to the natural isolation of the surface oxidation if the distance (L) between the inner edges of the two voltage electrodes were kept much larger than 1.0 μm. In our devices, the distance of two inner voltage electrodes is kept more than 2 μm.

## III. Detailed subtraction of the background for sample NR1

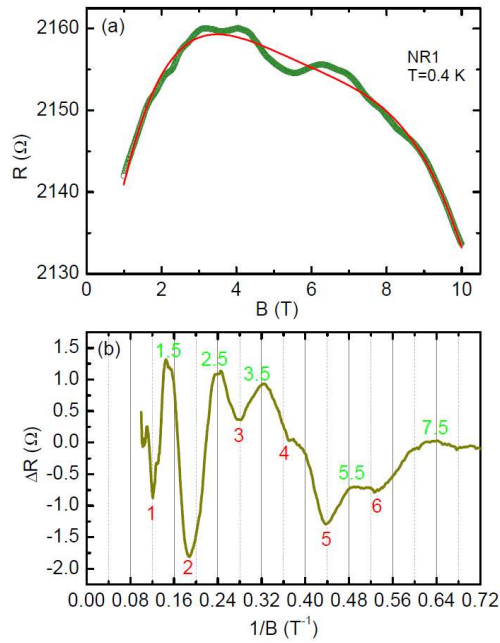

**Figure S2**. Background subtracted signal for low-field region MR curve of sample NR1 measured at 0.4 K with $B$ applied perpendicular to the nanoribbon ($\theta=0°$). (a) The open circle is the experimental data. The red solid curve is the fitting of the background with a smooth 4th order polynomial. (b) Amplitude of the resistance oscillations, $\Delta R$, in low $B$-regime. The marked numbers are the Landau levels for both the dips and peaks.

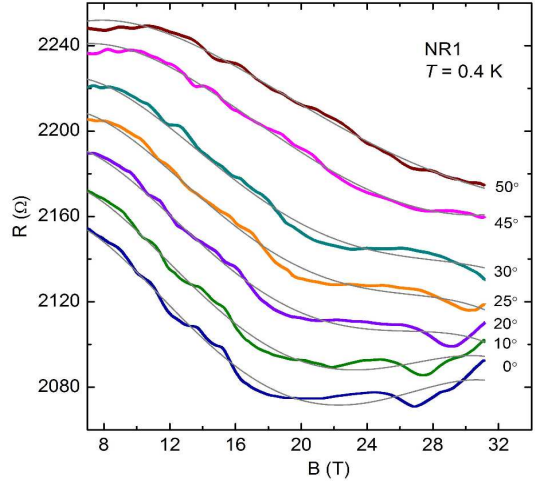

**Figure S3.** R versus B with various tilted angles measured at 0.4 K for sample NR1 in high field region, and the smooth line in each curve is the fit with 4th order polynomial function. The data at different tilted angles were offset for clarity.

## IV. The schematic of the Fermi surface of the hole and electron pockets in three planes of bismuth

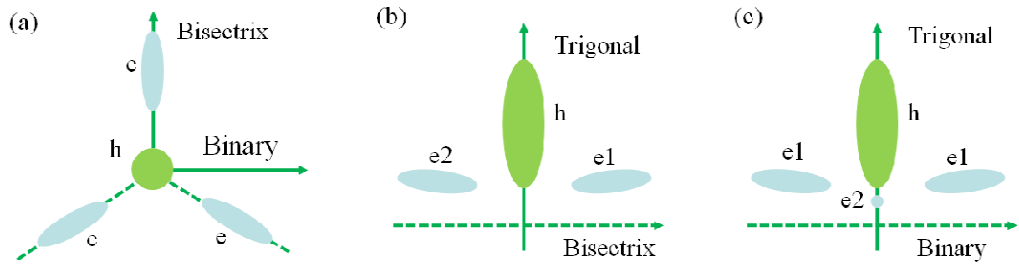

**Figure S4.** The extremal cross sections of the Fermi surface perpendicular to *B* are shown in the (a) B⊥current, B // nanoribbon, and B // trigonal axis; (b) B⊥current, B⊥nanoribbon, and B⊥trigonal axis; c) B // current, B//nanoribbon, and B⊥trigonal axis.[1]

## V. Temperature-dependent SdH oscillations for NR1 in the low-field regime and NR3

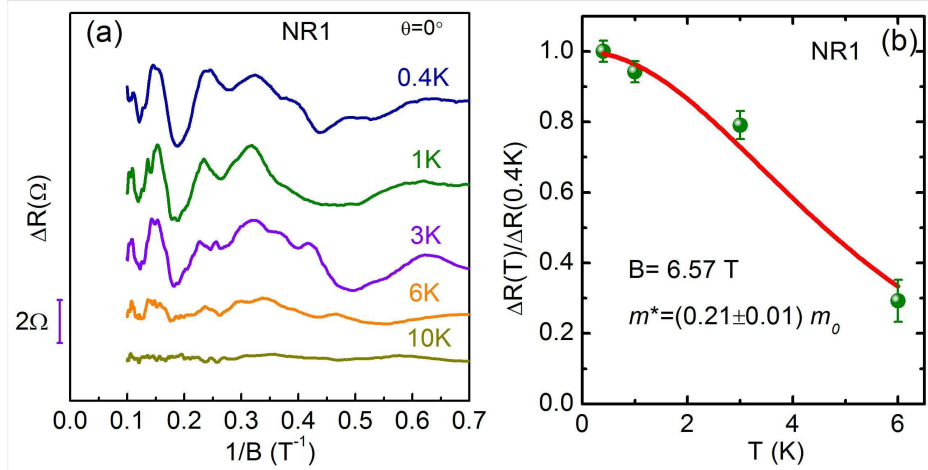

**Figure S5.** Temperature-dependent SdH oscillations for sample NR1 in the low-field regime. (**a**) The amplitude of the SdH oscillations versus $1/B$ under a perpendicular $B$ at different $T$. (b) Temperature dependence of the scaled oscillation amplitudes at $B=6.57$ T. From the fitting, the effective mass is estimated to be 0.21 $m_0$.

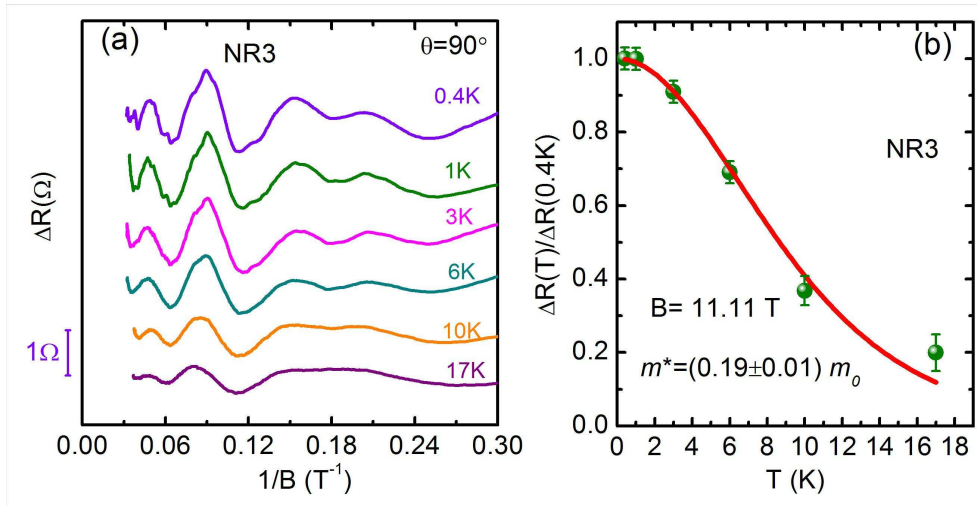

**Figure S6**. Temperature-dependent SdH oscillations for sample NR3. (**a**) The amplitude of the SdH oscillations versus $1/B$ under a perpendicular $B$ at different $T$. (b) Temperature dependence of the scaled oscillation amplitudes at $B=11.11$T. From the fitting, the effective mass is estimated to be 0.19 $m_0$.